%% file: main.tex
\documentclass{article}
\usepackage{authblk,amsmath,amssymb,geometry,graphicx,tikz,stmaryrd,inconsolata}
\usepackage[draft]{minted}
\usetikzlibrary{positioning}
\graphicspath{{images/}}

\title{Differentiable cellular automata}
\author{Carlos Martin}
\date{}
\affil{Columbia University}

\begin{document}

\maketitle

\begin{abstract}
    We describe a class of cellular automata (CAs) that are end-to-end differentiable. DCAs interpolate the behavior of ordinary CAs through rules that act on distributions of states. The gradient of a DCA with respect to its parameters can be computed with an iterative propagation scheme that uses previously-computed gradients and values. Gradient-based optimization over DCAs could be used to find ordinary CAs with desired properties.
\end{abstract}

\input{introduction.tex}
\input{ca.tex}
\input{probabilistic.tex}
\input{dca.tex}
\input{dca_examples.tex}
\input{dca_gradient.tex}
\input{dca_optimization.tex}
\input{bin_dca.tex}
\input{conclusion.tex}

\bibliographystyle{unsrt}
\bibliography{references}

\appendix
\input{code.tex}

\end{document}

%% file: introduction.tex
\section{Introduction}

A cellular automaton (CA) is a dynamical system consisting of a grid of cells, where each cell is in a particular state. At each timestep, the state of each cell is updated based on its current state and those of its neighbors. CAs can simulate physical \cite{Toffoli1984}\cite{Vichniac1984}\cite{Chopard2009}, chemical \cite{Lemont2005}, biological \cite{Hogeweg1988}\cite{Ermentrout1993}, and social \cite{Hegselmann1996}\cite{Nowak1996} processes. Some cellular automata are computationally universal \cite{Cook2009}\cite{Rendell2011}\cite{Rendell2016}.

An elementary cellular automaton (ECA) is a one-dimensional CA with two possible cell states, where the next state of a cell depends only on its current state and the states of its two immediate neighbors \cite{Wolfram2002}. An example of an ECA is rule 30, which has the following rule set:
\[ \LARGE
\substack{\blacksquare \blacksquare \blacksquare \\ \square} \enspace
\substack{\blacksquare \blacksquare \square \\ \square} \enspace
\substack{\blacksquare \square \blacksquare \\ \square} \enspace
\substack{\blacksquare \square \square \\ \blacksquare} \enspace
\substack{\square \blacksquare \blacksquare \\ \blacksquare} \enspace
\substack{\square \blacksquare \square \\ \blacksquare} \enspace
\substack{\square \square \blacksquare \\ \blacksquare} \enspace
\substack{\square \square \square \\ \square}
\]

The following diagram illustrates the evolution of this rule starting from a single black cell:
\begin{center}
\includegraphics[width=250pt]{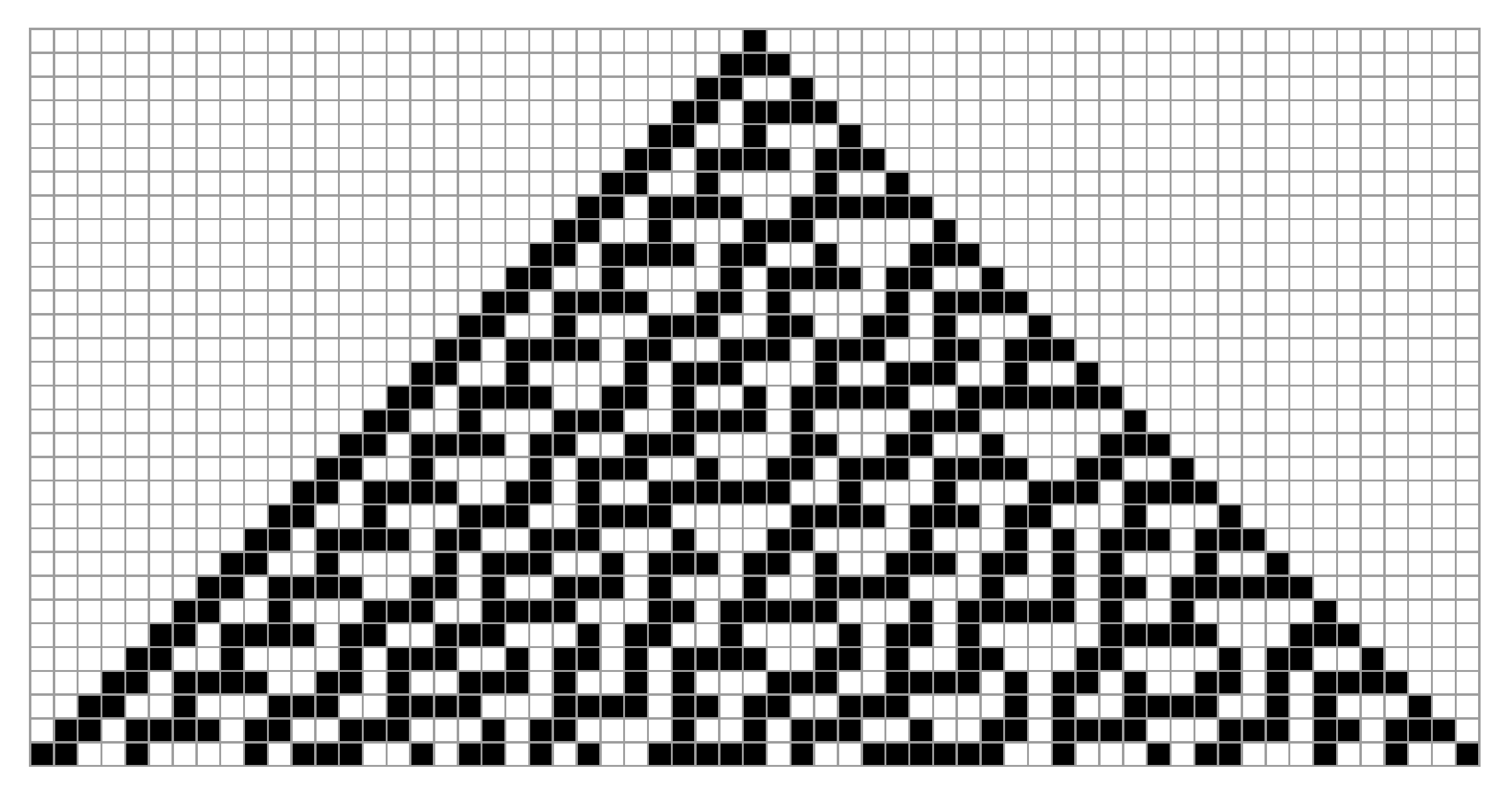}
\end{center}

Despite their simple rules, CAs exhibit a wide range of complex emergent behaviors \cite{Wolfram1983}\cite{Wolfram1984}\cite{Gutowitz1991}.

%% file: ca.tex
\section{Cellular automata}

A cellular automaton over a group \(G\) and alphabet \(A\) is a map \(\tau : A^G \rightarrow A^G\) such that
\[ \tau(x)(g) = \mu((x \circ L_g)|_S) \]

where \(S \subseteq G\), \(L_g : G \rightarrow G\) is the left multiplication by \(g\) in \(G\)
\[ L_g(g') = g g' \]

\(|_S : A^G \rightarrow A^S\) is the restriction from \(G\) to \(S\)
\[ x|_S(g) = x(g) \]

and \(\mu : A^S \rightarrow A\) \cite{Rka1999}\cite{CeccheriniSilberstein2010}\cite{Wacker2016}. \(S\) and \(\mu\) are the memory set and local map of the CA, respectively. For example, rule 30 has \(G = \mathbb{Z}/n\mathbb{Z}\), \(A = \{\square, \blacksquare\}\), \(S = \{-1, 0, 1\}\), and
\begin{align*}
    &\mu(\blacksquare\blacksquare\blacksquare) = \square
    &&\mu(\blacksquare\blacksquare\square) = \square
    &\mu(\blacksquare\square\blacksquare) = \square
    &&\mu(\blacksquare\square\square) = \blacksquare \\
    &\mu(\square\blacksquare\blacksquare) = \blacksquare
    &&\mu(\square\blacksquare\square) = \blacksquare
    &\mu(\square\square\blacksquare) = \blacksquare
    &&\mu(\square\square\square) = \square
\end{align*}

%% file: probabilistic.tex
\subsection{Probabilistic cellular automata}

Let \(\triangle A\) be the set of probability measures on \(A\):
\[ \triangle A = \left\{ P : A \rightarrow [0, 1] \;\middle|\; \sum_{a \in A} P(a) = 1 \right\} \]

In a probabilistic cellular automaton, the new states are sampled from a neighborhood-dependent probability distribution \(\mu : A^S \rightarrow \triangle A\). They are described in detail in \cite{Grinstein1985}\cite{Lebowitz1990}\cite{Schle2009}\cite{Bui2013}. For example, suppose we have a probability distribution \(\triangle (A^S)\) over possible neighborhoods and wish to find the probability distribution for the new state of the cell. Then for \(X \in \triangle (A^S)\),
\begin{align*}
    \mathrm{P}(\mu(X) = a)
    &= \sum_{x \in A^S} \mathrm{P}(\mu(X) = a \mid X = x) \mathrm{P}(X = x) \\
    &= \sum_{x \in A^S} \mathrm{P}(\mu(x) = a) \mathrm{P}(X = x)
\end{align*}

As an approximation, we assume the states of distinct cells in the neighborhood are independent:
\[ \forall s,s' \in S : s \neq s' \rightarrow X(s) \perp X(s') \]

This allows us to factor the neighborhood distribution into the individual state distributions:
\[ \mathrm{P}(X = x) = \prod_{s \in S} \mathrm{P}(X(s) = x(s)) \]

Hence
\[ \mathrm{P}(\mu(X) = a) = \sum_{x \in A^S} \mathrm{P}(\mu(x) = a) \prod_{s \in S} \mathrm{P}(X(s) = x(s)) \]

The independence assumption implies \(X\) can be described as an element of \((\triangle A)^S\).

%% file: dca.tex
\section{Differentiable cellular automata}

A differentiable cellular automaton (DCA) over a group \(G\) and alphabet \(A\) is a cellular automaton over \(G\) and \(\triangle A\) such that

\[ \mu(x)(a) = \sum_{y \in A^S} \rho(y)(a) \prod_{s \in S} x(s)(y(s)) \]

where \(\rho : A^S \rightarrow \triangle A\). It behaves like an ordinary cellular automaton over \(G\) and \(A\) when \(\rho\) yields deterministic distributions:
\[ \forall y \in A^S : \exists a \in A : \rho(y) = \delta(a) \]

where \(\delta\) is the discrete delta function
\[ \delta(a)(a') = \begin{cases}
    1 & a = a' \\
    0 & \text{otherwise}
\end{cases}\]

Otherwise, it behaves like a mixture of ordinary cellular automata over \(G\) and \(A\). Let \(\rho = \sigma \circ w\) where \(\sigma : (A \rightarrow \mathbb{R}) \rightarrow \triangle A\) is the softmax function
\[ \sigma(z)(a) = \frac{\exp z(a)}{\sum_{a' \in A} \exp z(a')} \]

and \(w : A^S \rightarrow A \rightarrow \mathbb{R}\) assigns a real-valued weight to each pair \((y, a) \in A^S \times A\). It is sometimes convenient to parameterize with respect to \(w\) rather than \(\rho\) because \(\sigma \circ w\) is always normalized.

%% file: dca_examples.tex
\subsection{Examples}

Consider two ECA rules that differ only in the output for \(\blacksquare\blacksquare\square\):
\begin{align*}
\LARGE
\substack{\blacksquare \blacksquare \blacksquare \\ \square} \enspace
\substack{\blacksquare \blacksquare \square \\ ?} \enspace
\substack{\blacksquare \square \blacksquare \\ \square} \enspace
\substack{\blacksquare \square \square \\ \blacksquare} \enspace
\substack{\square \blacksquare \blacksquare \\ \blacksquare} \enspace
\substack{\square \blacksquare \square \\ \blacksquare} \enspace
\substack{\square \square \blacksquare \\ \blacksquare} \enspace
\substack{\square \square \square \\ \square}
\end{align*}

The following diagrams illustrate their evolution starting from the same configuration:
\begin{center}
\includegraphics[width=150pt]{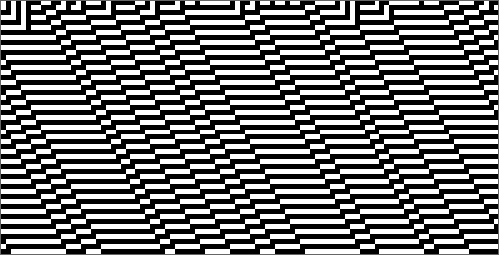} \quad 
\includegraphics[width=150pt]{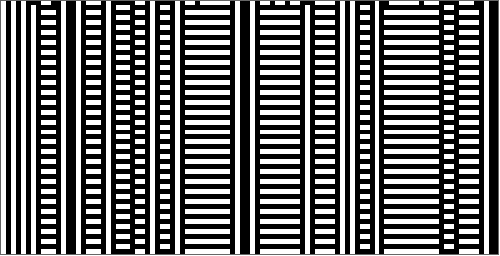}
\end{center}

A DCA that interpolates these two CAs is given by
\begin{align*}
    &\rho(\blacksquare\blacksquare\blacksquare)(\blacksquare) = 0
    &&\rho(\blacksquare\blacksquare\square)(\blacksquare) = \alpha
    &\rho(\blacksquare\square\blacksquare)(\blacksquare) = 0
    &&\rho(\blacksquare\square\square)(\blacksquare) = 1 \\
    &\rho(\square\blacksquare\blacksquare)(\blacksquare) = 1
    &&\rho(\square\blacksquare\square)(\blacksquare) = 1
    &\rho(\square\square\blacksquare)(\blacksquare) = 1
    &&\rho(\square\square\square)(\blacksquare) = 1
\end{align*}

where \(\alpha \in [0, 1]\). \(\alpha = 0\) and \(\alpha = 1\) yield the CA on the left and right, respectively. For \(0 < \alpha < 1\), starting with the same configuration, we obtain the following diagrams:

\begin{center}
\includegraphics[width=130pt]{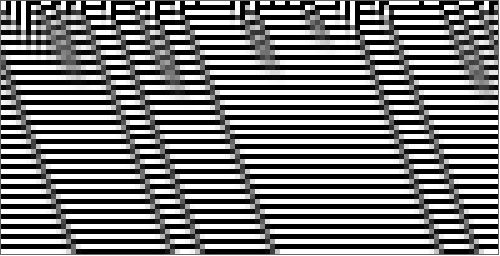}
\includegraphics[width=130pt]{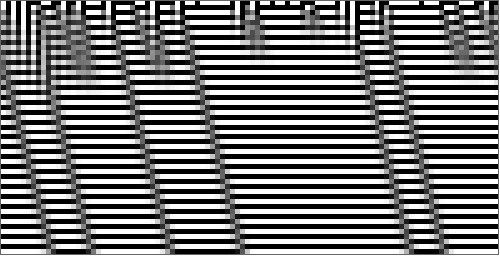}
\includegraphics[width=130pt]{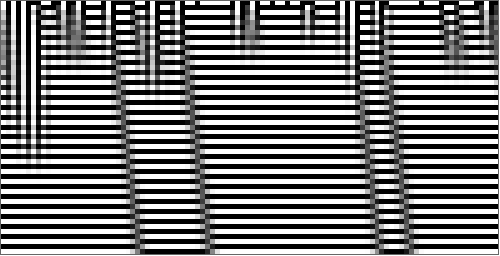}
\end{center}

where a grayscale is used to indicate the probability of \(\blacksquare\). Notice their behavior is ``in-between'' those of the two original CAs. Similarly,
\begin{align*}
    &\rho(\blacksquare\blacksquare\blacksquare)(\blacksquare) = 1
    &&\rho(\blacksquare\blacksquare\square)(\blacksquare) = 0
    &\rho(\blacksquare\square\blacksquare)(\blacksquare) = 1
    &&\rho(\blacksquare\square\square)(\blacksquare) = 0 \\
    &\rho(\square\blacksquare\blacksquare)(\blacksquare) = 1
    &&\rho(\square\blacksquare\square)(\blacksquare) = 1
    &\rho(\square\square\blacksquare)(\blacksquare) = \alpha
    &&\rho(\square\square\square)(\blacksquare) = 0
\end{align*}

for \(\alpha \in \{0, .2, .4, .6, .8, 1\}\) yields the following diagrams in clockwise order:

\begin{center}
\includegraphics[width=130pt]{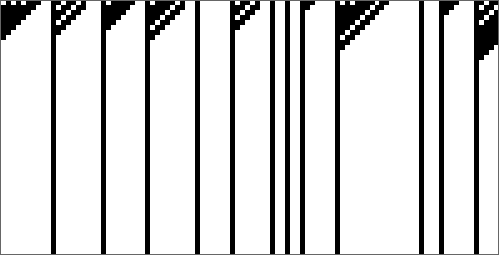}
\includegraphics[width=130pt]{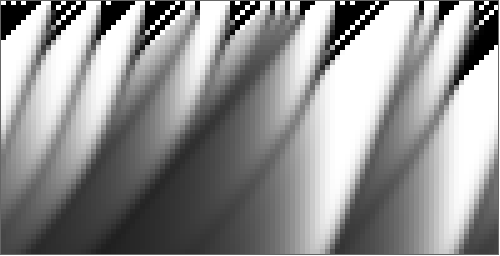}
\includegraphics[width=130pt]{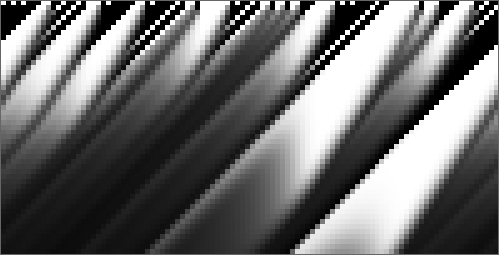} \\
\includegraphics[width=130pt]{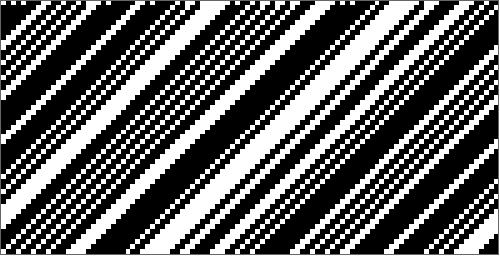}
\includegraphics[width=130pt]{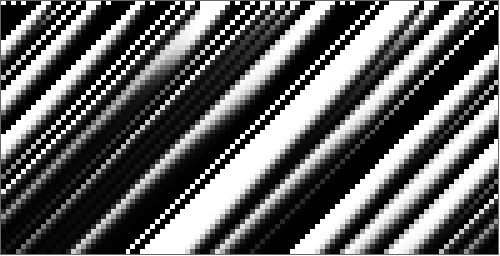}
\includegraphics[width=130pt]{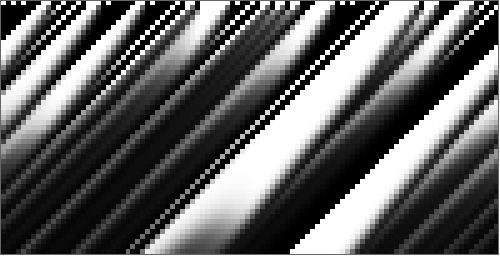}
\end{center}

%% file: dca_gradient.tex
\subsection{Gradient}

The derivative of \(\mu(x)(a)\) with respect to a weight \(w(y')(a')\) is
\begin{align*}
    \frac{\partial \mu(x)(a)}{\partial w(y')(a')}
    &= \sum_{y \in A^S} \frac{\partial}{\partial w(y')(a')}  \rho(y)(a) \prod_{s \in S} x(s)(y(s)) \\
    &= \sum_{y \in A^S} \left( \frac{\partial \rho(y)(a)}{\partial w(y')(a')} \prod_{s \in S} x(s)(y(s)) + \rho(y)(a) \frac{\partial}{\partial w(y')(a')} \prod_{s \in S} x(s)(y(s)) \right) \\
    &= \frac{\partial \rho(y')(a)}{\partial w(y')(a')} \prod_{s \in S} x(s)(y'(s)) + \sum_{y \in A^S} \rho(y)(a) \frac{\partial}{\partial w(y')(a')} \prod_{s \in S} x(s)(y(s)) \\
    &= \frac{\partial \sigma(w(y'))(a)}{\partial w(y')(a')} \prod_{s \in S} x(s)(y'(s)) + \sum_{y \in A^S} \rho(y)(a) \frac{\partial}{\partial w(y')(a')} \prod_{s \in S} x(s)(y(s))
\end{align*}

The derivative of the softmax function is
\[ \frac{\partial \sigma(z)(a)}{\partial z(a')} = \sigma(z)(a) (\delta(a)(a') - \sigma(z)(a')) \]

Hence
{\footnotesize \begin{align*}
    \frac{\partial \mu(x)(a)}{\partial w(y')(a')}
    &= \rho(y')(a) (\delta(a)(a') - \rho(y')(a')) \prod_{s \in S} x(s)(y'(s)) + \sum_{y \in A^S} \rho(y)(a) \frac{\partial}{\partial w(y')(a')} \prod_{s \in S} x(s)(y(s)) \\
    &= \rho(y')(a) (\delta(a)(a') - \rho(y')(a')) \prod_{s \in S} x(s)(y'(s)) + \sum_{y \in A^S} \rho(y)(a) \sum_{s \in S} \frac{\partial x(s)(y(s))}{\partial w(y')(a')} \prod_{s' \in S \setminus \{s\}} x(s')(y(s'))
\end{align*}}

which implies
{\footnotesize \[
    \frac{\partial \varphi(x)(g)(a)}{\partial w(y')(a')}
    = \rho(y')(a) (\delta(a)(a') - \rho(y')(a')) \prod_{s \in S} x(g s)(y'(s)) + \sum_{y \in A^S} \rho(y)(a) \sum_{s \in S} \frac{\partial x(g s)(y(s))}{\partial w(y')(a')} \prod_{s' \in S \setminus \{s\}} x(g s')(y(s'))
\]}

Thus the derivative of the new state distribution of a cell with respect to a weight depends on the neighboring state distributions as well as \textit{their} derivatives with respect to this weight. Hence we can compute the weight derivatives of a new configuration from the old configuration and its derivatives. This is illustrated in the following diagram, where the gradient of a configuration \(\tau^k(x)\) is taken with respect to \(w(y)(a)\) for every \(y \in A^H\) and \(a \in A\):
\begin{center}
\begin{tikzpicture}[node distance=1cm and 2cm]
    \node (a) {configurations};
    \node [right=of a] (b) {configuration gradients};

    \node [below=.25cm of a] (a0) {\(x\)};
    \node [below=.25cm of b] (b0) {\(\nabla x = 0\)};
    
    \node [below=of a0] (a1) {\(\tau(x)\)};
    \node [below=of b0] (b1) {\(\nabla \tau(x)\)};
    
    \node [below=of a1] (a2) {\(\tau^2(x)\)};
    \node [below=of b1] (b2) {\(\nabla \tau^2(x)\)};
    
    \node [below=of a2] (a3) {\(\tau^3(x)\)};
    \node [below=of b2] (b3) {\(\nabla \tau^3(x)\)};
    
    \draw [->]
        (a0) edge (a1) 
        (a1) edge (a2) 
        (a2) edge (a3)
        
        (b0) edge (b1)
        (b1) edge (b2)
        (b2) edge (b3)
        
        (a0) edge (b1)
        (a1) edge (b2)
        (a2) edge (b3)
    ;
\end{tikzpicture}
\end{center}

The arrows indicate the direction of dependencies between computations.

%% file: dca_optimization.tex
\subsection{Optimization}

Suppose we want to have \(\tau^n(x) = \varphi(x)\) for some \(x \in A^G\), \(n \in \mathbb{N}\), and \(\varphi : A^G \rightarrow A^G\). \(\varphi\), for example, could be the following ``majority'' function:
\[ \varphi(x)(g) = \delta\left( \operatorname*{argmax}_{a \in A} \sum_{g' \in G} x(g')(a) \right) \]

Then we could try to minimize
\[ E = \sum_{g \in G} \mathrm{H}(\varphi(x)(g), \tau^n(x)(g)) \]

where \(\mathrm{H}\) is the cross entropy between the target distribution \(p\) and current distribution \(\hat{p}\):
\[ \mathrm{H}(p, \hat{p}) = -\sum_{\omega \in \Omega} p(\omega) \log \hat{p}(\omega) \]

Thus
\[ E = -\sum_{g \in G} \sum_{a \in A} \varphi(x)(g)(a) \log \tau^n(x)(g)(a) \]

Taking the gradient yields
\[ \nabla E = -\sum_{g \in G} \sum_{a \in A} \frac{\varphi(x)(g)(a)}{\tau^n(x)(g)(a)} \nabla \tau^n(x)(g)(a) \]

where \(\nabla \tau^n(x)\) is computed using the procedure described in the previous section. This gradient allows us to adjust the weights \(w\) to minimize \(E\). For example, we could iterate
\begin{align*}
    w_{i+1} &= w_i - \varepsilon \nabla E(w_i)
\end{align*}

where \(\varepsilon > 0\) is a descent rate. More sophisticated optimization techniques can also be used. If our goal is to have \(\tau^n = f\) in general, we sum over all possible initial configurations:
\[ E = \sum_{x \in A^G} \sum_{g \in G} \mathrm{H}(\varphi(x)(g), \tau^n(x)(g)) \]

This can be approximated by summing over a proper subset of \(A^G\) instead.

%% file: bin_dca.tex
\section{The binary case}

A binary DCA is equivalent to an ordinary cellular automaton with alphabet \([0,1]\), where the latter represents the probability of being in one of the two states. Then
\begin{align*}
\mu(x)
&= \sum_{y \in \{0,1\}^S} \rho(y) \prod_{s \in S} \begin{cases}
    x(s) & y(s) = 1 \\
    1 - x(s) & \text{otherwise}
\end{cases} \\
&= \sum_{y \in \{0,1\}^S} \rho(y) \prod_{s \in S} \langle x(s), y(s) \rangle
\end{align*}

where \(\rho = \sigma \circ w\), \(w : \{0,1\}^S \rightarrow \mathbb{R}\), \(\sigma : \mathbb{R} \rightarrow [0, 1]\) is the sigmoid function, and

\[\langle x(s), y(s) \rangle = x(s) y(s) + (1 - x(s)) (1 - y(s))\]

Its gradient is
\[
    \frac{\partial \mu(x)}{\partial w(y')}
    = \rho(y') (1 - \rho(y')) \prod_{s \in S} \langle x(s), y'(s) \rangle + \sum_{y \in \{0,1\}^S} \rho(y) \sum_{s \in S} \frac{\partial \langle x(s), y(s) \rangle}{\partial w(y')} \prod_{s' \in S \setminus \{s\}} \langle x(s'), y(s') \rangle \\
\]

Note that
\begin{align*}
\frac{\partial \langle x(s), y(s) \rangle}{\partial w(y')} 
&= \frac{\partial x(s)}{\partial w(y')} y(s) + \frac{\partial (1 - x(s))}{\partial w(y')} (1 - y(s)) \\
&= \frac{\partial x(s)}{\partial w(y')} (2 y(s) - 1)
\end{align*}

The corresponding error function is given by
\[ E = -\sum_{g \in G} \Big( \varphi(x)(g) \log \tau^n(x)(g) + (1 - \varphi(x)(g)) \log (1 - \tau^n(x)(g)) \Big) \]

Its gradient is
\[ \nabla E = -\sum_{g \in G} \left( \frac{\varphi(x)(g)}{\tau^n(x)(g)} - \frac{1 - \varphi(x)(g)}{1 - \tau^n(x)(g)} \right) \nabla \tau^n(x)(g) \]

A code example for simulating binary DCAs is included in the appendix.

%% file: conclusion.tex
\section{Conclusion}

In this paper, we have described a class of CAs that are end-to-end differentiable. DCAs interpolate the behavior of ordinary CAs through rules that act on distributions of states rather single states. The gradient of a DCA with respect to its parameters can be computed with an iterative propagation scheme that uses previously-computed gradients and values.

Representing the fitness of a DCA rule with a differentiable loss function allows gradient-based global optimization techniques to be used to speed up search. Candidates include gradient-informed simulated annealing \cite{Yiu2004}, gradient tabu search \cite{Stepanenko2006}, function stretching techniques \cite{Wang2007}, and gradient-based cuckoo search \cite{Fateen2014}. We hope to see these methods applied to DCAs in future research, allowing the exponentially-large space of CA rules to be searched more efficiently.

%% file: code.tex
\section{Example code}

The following Python code contains functions for computing values and gradients of a differentiable cellular automata. It includes a comparison of the error gradient obtained with the scheme described in the paper and that obtained with a finite-difference approximation.

\begin{minted}{python}
import numpy as np
import matplotlib.pyplot as plt
import scipy.special
import itertools
import scipy.ndimage

np.set_printoptions(linewidth=np.inf)

def local_map(neighborhood, patterns, outputs):
  return np.sum(outputs * np.prod(neighborhood * patterns + (1 - neighborhood) * (1 - patterns), axis=1), axis=0)

def global_map(configuration, patterns, outputs):
  return scipy.ndimage.generic_filter(configuration, local_map, size=patterns.shape[1], mode='wrap',
    extra_arguments=(patterns, outputs))

def local_grad_map(neighborhood, neighborhood_gradient, patterns, outputs):
  return outputs * (1 - outputs) * np.prod(patterns * neighborhood + (1 - patterns) * (1 - neighborhood), axis=1) + \
    np.sum(outputs * np.prod(patterns * neighborhood + (1 - patterns) * (1 - neighborhood), axis=1) * np.sum( (np.rollaxis(np.tile(neighborhood_gradient, (outputs.shape[0], 1, 1)), 2) * (2*patterns - 1) / (patterns * neighborhood + (1 - patterns) * (1 - neighborhood))) , axis=2) , axis=1)

def global_grad_map(configuration, configuration_gradient, patterns, outputs):
  neighbors_centered = range(-patterns.shape[1]//2 + 1, patterns.shape[1]//2 + 1)
  return np.array([local_grad_map(a_, b_, patterns, outputs) for a_, b_ in zip(
    np.transpose([ np.roll(configuration, i) for i in neighbors_centered ]), 
    np.transpose([ np.roll(configuration_gradient, i) for i in neighbors_centered ], (1, 0, 2))
  )])

def evaluate(initial, depth, patterns, outputs):
  configuration = initial
  configuration_gradient = np.zeros((initial.shape[0], outputs.shape[0]))
  for layer in range(depth):
    configuration, configuration_gradient = global_map(configuration, patterns, outputs), global_grad_map(configuration, configuration_gradient, patterns, outputs)
  return configuration, configuration_gradient

def evaluate_diagram(initial, depth, patterns, outputs):
  diagram = np.ndarray((depth, initial.shape[0]))
  diagram_gradient = np.ndarray((depth, initial.shape[0], outputs.shape[0]))
  diagram[0] = initial
  diagram_gradient[0] = np.zeros((initial.shape[0], outputs.shape[0]))
  for layer in range(depth - 1):
    diagram[layer + 1] = global_map(diagram[layer], patterns, outputs)
    diagram_gradient[layer + 1] = global_grad_map(diagram[layer], diagram_gradient[layer], patterns, outputs)
  return diagram, diagram_gradient

def loss_function(weights, patterns, initial, depth, target_function):
  outputs = scipy.special.expit(weights)
  target = target_function(initial)
  configuration, configuration_gradient = evaluate(initial, depth, patterns, outputs)
  error = -np.sum(target * np.log(configuration) + (1 - target) * np.log((1 - configuration)))
  error_gradient = -np.sum(configuration_gradient.T * (target / configuration - (1 - target) / (1 - configuration)), axis=1)
  return error / initial.shape[0], error_gradient / initial.shape[0]

def batch_loss_function(weights, patterns, initials, depth, target_function):
  batch_error = 0
  batch_error_gradient = np.zeros(weights.shape)
  for initial in initials:
    error, error_gradient = loss_function(weights, patterns, initial, depth, target_function)
    batch_error += error
    batch_error_gradient += error_gradient
  return batch_error / len(initials), batch_error_gradient / len(initials)

def plot_example(size=500, depth=300):
  patterns = np.array(tuple(itertools.product([0, 1], repeat=3)))
  initial = np.random.choice([1e-3,1 - 1e-3], size=size)
  outputs = np.array([0, .5, 1, 1, 0, 1, 0, 1])

  diagram, diagram_gradient = evaluate_diagram(initial, depth, patterns, outputs)

  fig, axes = plt.subplots(nrows=4, ncols=2)
  for i in range(outputs.shape[0]):
    axes[i // 2][i % 2].set_title('Gradients wrt weight of {0:b}'.format(i).zfill(3))
    axes[i // 2][i % 2].invert_yaxis()
    axes[i // 2][i % 2].pcolormesh(diagram_gradient[:,:,i], cmap='gray_r')
    axes[i // 2][i % 2].tick_params(axis='x', which='both', bottom='off', top='off', labelbottom='off')
    axes[i // 2][i % 2].get_xaxis().set_visible(False)
    axes[i // 2][i % 2].get_yaxis().set_visible(False)
  fig = plt.figure()
  fig.gca().set_title('Values')
  fig.gca().invert_yaxis()
  fig.gca().pcolormesh(diagram, cmap='gray_r', vmin=0, vmax=1)
  fig.gca().tick_params(axis='x', which='both', bottom='off', top='off', labelbottom='off')
  fig.gca().get_xaxis().set_visible(False)
  fig.gca().get_yaxis().set_visible(False)
  plt.axes().set_aspect('equal', 'datalim')
  plt.show()

def test_gradients(radius=2, size=100, depth=20, target_function=lambda configuration: configuration):
  print('Testing accuracy of error gradient...\n')
  patterns = np.array(tuple(itertools.product([0, 1], repeat=2*radius+1)))
  for iteration in itertools.count():
    weights = np.random.normal(size=2**(2*radius+1))
    initial = np.random.choice([1e-3,1 - 1e-3], size=size)
    
    f = lambda w: loss_function(w, patterns, initial, depth, target_function)[0]
    fprime = lambda w: loss_function(w, patterns, initial, depth, target_function)[1]
    print('Propagation scheme:\t{}'.format(fprime(weights)))
    print('Finite-difference:\t{}'.format(scipy.optimize.approx_fprime(weights, f, 1e-8)))
    print('2-norm of difference:\t{}\n'.format(scipy.optimize.check_grad(f, fprime, weights)))

def irprop_minus(func, param, args, init_step_size=.0125, eta_plus=1.2, eta_minus=.5, min_step_size=0, max_step_size=50):
  param = param.copy()
  step_sizes = np.full(param.shape, init_step_size)
  prev_error_gradient = np.zeros(param.shape)
  
  for iteration in itertools.count():
    error, error_gradient = func(param, *args)
    yield error, error_gradient, param
    step_sizes[error_gradient * prev_error_gradient > 0] *= eta_plus
    step_sizes[error_gradient * prev_error_gradient < 0] *= eta_minus
    np.clip(step_sizes, min_step_size, max_step_size, step_sizes)

    param -= np.sign(error_gradient) * step_sizes
    prev_error_gradient = error_gradient

def irprop_plus(func, param, args, init_step_size=.0125, eta_plus=1.2, eta_minus=.5, min_step_size=0, max_step_size=50):
  param = param.copy()
  step_sizes = np.full(param.shape, init_step_size)
  prev_error_gradient = np.zeros(param.shape)
  
  arg_change = np.zeros(param.shape)
  prev_error = 0

  for iteration in itertools.count():
    error, error_gradient = func(param, *args)
    yield error, error_gradient, param
    step_sizes[error_gradient * prev_error_gradient > 0] *= eta_plus
    step_sizes[error_gradient * prev_error_gradient < 0] *= eta_minus
    np.clip(step_sizes, min_step_size, max_step_size, step_sizes)

    arg_change = np.where(
      error_gradient * prev_error_gradient >= 0,
      -np.sign(error_gradient) * step_sizes,
      -arg_change if error > prev_error else 0
    )
    param += arg_change
    prev_error = error
    prev_error_gradient = np.where(
      error_gradient * prev_error_gradient >= 0,
      error_gradient,
      0
    )

def descent_example(radius=2, size=100, depth=20, target_function=lambda configuration: configuration):
  patterns = np.array(tuple(itertools.product([0, 1], repeat=2*radius+1)))

  initials = [np.random.choice([1e-3,1 - 1e-3], size=size) for i in range(10)]

  plt.ion()
  plt.xlabel('Iteration')
  plt.ylabel('Average cross entropy')
  plt.title('{} neighbors, {} cells, {} layers'.format(2*radius+1, size, depth))
  line, = plt.plot([])

  weights = np.random.normal(size=2**(2*radius+1))

  for iteration, (error, error_gradient, weights) in enumerate(irprop_plus(batch_loss_function, weights, (patterns, initials, depth, target_function))):
    print('error: {:.4f}'.format(error))
    print('neighborhood\tprobability\terror gradient')
    print('\n'.join('{}\t{:.4f}\t\t{:.4f}'.format(template, template_probability, template_error_gradient) for template, template_probability, template_error_gradient in zip(patterns, scipy.special.expit(weights), error_gradient)))
    print('')

    line.set_xdata(np.append(line.get_xdata(), [iteration]))
    line.set_ydata(np.append(line.get_ydata(), [error]))

    plt.gca().relim()
    plt.gca().autoscale_view()
    plt.pause(.001)

plot_example()
test_gradients()
\end{minted}